\documentclass[letterpaper, 10 pt, conference]{ieeeconf}  % Comment this line out if you need a4paper
\IEEEoverridecommandlockouts                             
\usepackage{graphicx}
\usepackage{subcaption} 
\usepackage{hyperref} 
\usepackage{epsfig} 
\usepackage{times} 
\usepackage{amsmath,amssymb,cite} 
\usepackage{mathrsfs} 
\usepackage{booktabs}
\usepackage{algorithmic}
\usepackage{algorithm}
\graphicspath{{Figs/}}
\title{\LARGE \bf
Enhancing Mobile Robot Navigation Safety and Efficiency through NMPC with Relaxed CBF in Dynamic Environments
}
\author{Nhat Nguyen Minh$^{1}$, Stephen McIlvanna$^{1}$, Yuzhu Sun$^{1}$, Yan Jin$^{2}$, and Mien Van$^{1}$
\thanks{$^{1}$School of Electronics, Electrical Engineering and Computer Science, Queen’s University Belfast, Belfast, United Kingdom
        {\tt\small m.van@qub.ac.uk}}%
\thanks{$^{2}$School of Mechanical and Aerospace Engineering, Queen’s University Belfast, Belfast, United Kingdom}
}
\begin{document}
\maketitle
\thispagestyle{empty}
\pagestyle{empty}
%%%%%%%%%%%%%%%%%%%%%%%%%%%%%%%%%%%%%%%%%%%%%%%%%%%%%%%%%%
\begin{abstract}
In this paper, a safety-critical control strategy for a nonholonomic robot is developed to generate control signals that result in optimal, obstacle-free paths through dynamic environments. We formulate the control synthesis problem as an Optimal Control Problem (OCP) that enforces Control Lyapunov Function (CLF) constraints for system stability as well as safety-critical constraints using Control Barrier Function (CBF) with a relaxing decay rate of the barrier function. A Nonlinear Model Predictive Control (NMPC) integrates with CLF and CBF to ensure system safety and facilitate optimal performance within a short prediction horizon, reducing the computational burden in real-time implementation. Additionally, we incorporate an obstacle avoidance constraint based on the Euclidean norm into the NMPC framework, showcasing the CBF approach's superiority in addressing mobile robotic systems' point stabilisation and trajectory tracking challenges. Through extensive simulations, the proposed controller demonstrates proficiency in static and dynamic obstacle avoidance under various scenarios. Experimental validations conducted using the Husky A200 robot align with simulation results, reinforcing the applicability of our proposed approach in real-world scenarios, notably improving the computational efficiency and safety in practical mobile robot applications.
\end{abstract}
%%%%%%%%%%%%%%%%%%%%%%%%%%%%%%%%%%%%%%%%%%%%%%%%%%%%%%%%%%
\section{INTRODUCTION}
Over the past few decades, there has been exponential growth in robotics and autonomous technologies, leading to a wide range of applications for wheeled mobile robots (WMRs) in industries such as logistics, discovery, search, and rescue \cite{ref1}. Considering the nonholonomic nature of WMRs, various control approaches have been devised to address point stabilisation. Studies tackling this problem include the use of differential kinematic controller \cite{ref6}, backstepping technique \cite{ref7}, and feedback linearisation \cite{ref8}. In the context of trajectory tracking, the literature presents numerous techniques, such as dynamic feedback linearisation method \cite{ref9}, backstepping techniques \cite{ref10}, and sliding mode technique \cite{ref12}. However, it is crucial to note that these controllers typically require the reference trajectories to remain persistently excited. Furthermore, the aforementioned methods are not optimization-based control strategies, which means they might not generate efficient control inputs that are subject to the system's physical constraints.
\\
To address these issues, we adopt NMPC, known for its ability to explicitly handle both soft and hard constraints and for its capacity for simultaneous tracking and regulation. NMPC formulates an online optimisation problem that predicts system outputs based on current states and system models over a predefined prediction horizon, all while maintaining compliance with a set of control actions and state constraints. Various MPC variants, including linear MPC and nonlinear MPC, have been employed to address point stabilisation and tracking issues in mobile robots \cite{ref2}, \cite{ref15}.
\\
Since mobile robots operate autonomously in dynamic environments, including varying references and obstacles, safety and stability are crucial factors. Robotic navigation approaches can be categorised into three main types: reactive-based, learning-based, and optimization-based. Reactive methods, such as those by \cite{n1} and \cite{n2}, focus on real-time collision avoidance. Learning-based strategies, like those by \cite{n3} and \cite{n4}, use deep reinforcement learning to navigate complex environments. Optimization-based techniques, discussed by \cite{n5} and \cite{n6}, employ advanced algorithms such as NMPC for optimal path planning and collision avoidance in dynamic settings.
\\
As outlined in \cite{r2.6}, the CBF approach provides a structured framework extensively used to implement safety constraints like collision avoidance in robotic systems. The decay rate of the barrier function, $\gamma$, is a tunable parameter that plays a vital role in determining the system's adherence to safety constraints. \cite{r2.9} has proposed a new CBF with a relaxed $\gamma$ technique that can enhance both system safety and performance. On the other hand, \cite{r2.8} demonstrated the improved performance and computational time of NMPC with constraint-based CLF on a robot platform. 
\\
Motivated by the considerations outlined above, this work proposes a Nonlinear Model Predictive Control (NMPC) scheme, integrating a CLF and a CBF with a relaxed decay rate across the prediction horizon for mobile robots. Multiple challenging scenarios involving static and dynamic obstacles are devised in MATLAB to evaluate the safety performance of the NMPC-CLF-Relax-CBF approach. A subsequent comparative analysis among the Relax-CBF, traditional CBF, and the Bug-Type (BT) algorithm emphasises the use of a short prediction time and a high decay rate of barrier function. Moreover, the practical applicability of the proposed scheme is validated for the first time on a physical robot platform using the Husky A200. 
\\
The remainder of the paper is organised as follows: Section II provides the theoretical background on the mobile robot's kinematic model. Section III details the NMPC design and implementation, including the discrete-time CLF and CBF with a relaxed decay rate. Section IV presents simulation and experimental results in safety-critical autonomous driving scenarios. Lastly, Section V concludes our work.
\section{MATHEMATICAL MODELING}
The state of the nonholonomic mobile robot is defined by a state vector $\xi = [x, y, \theta]^T \in \Re^3$, which includes the spatial position $\xi_{xy} = [x, y[^T$ (m,m) and the orientation angle $\theta$ (rad) of the robot. The robot's control inputs are represented by a vector $u = [v, w]^T \in \Re^2$, where $v$ (m/s) and $w$ (rad/s) correspond to linear velocity and angular velocity, respectively. This study considers the kinematic model of a nonholonomic mobile robot as follows:
\begin{equation} \label{eq1}
\dot{\xi}(t) =f(\xi(t),u(t)) =
	\left[ \begin{matrix}
		\cos(\theta(t))&0\\
		\sin(\theta(t))&0\\
		0&1\\
	\end{matrix} \right] 
	\left[ \begin{matrix}
		v(t)\\
		w(t)\\
	\end{matrix} \right]
\end{equation}
To derive the control problems, a reference robot is assumed to have the same constraints as (\ref{eq1}), and its kinematic model is written as:
\begin{equation} \label{eq2}
\dot{\xi}_{r}(t) =f(\xi_{r}(t),u_{r}(t)) =
	\left[ \begin{matrix}
		\cos(\theta_{r}(t))&0\\
		\sin(\theta_{r}(t))&0\\
		0&1\\
	\end{matrix} \right] 
	\left[ \begin{matrix}
		v_{r}(t)\\
		w_{r}(t)\\
	\end{matrix} \right] 
\end{equation}
The control challenge aims to minimise the difference between a robot's observed state and control inputs and those of a reference robot. To align the observed robot with the reference model, we use a discretised kinematic model with a time-step $\tau>0$. Denoting $\xi(k)=\xi(t_k)$, the integration over the fixed interval is numerically approximated with a piecewise constant control during each sampling interval and the Runge-Kutta4 method.
\begin{equation} \label{eq3}
\begin{aligned} 
\xi(k+1) &= \mathbf{F}( \xi(k),u(k))\\
         	     &=\xi (k)+\int_{t_k}^{t_k+\tau}{f(\xi (\varrho ),u(\varrho ))d\varrho}
\end{aligned}          
\end{equation}
where $k\in \mathbb{N}_0$ is the current sampling instant and $\mathbf{F}(\cdot): \Re^3\times \Re^2\rightarrow \Re^3$ is the discrete nonlinear dynamical mapping. In the point stabilisation control problem, the feedback control is designed such that the solution of (\ref{eq3}) starting from the feedback condition $\xi_{\text{fb}} \in \mathbf{X}$ stays close to a desired set point $\xi _{r}\in \mathbf{X}$, and converges, i.e.
\begin{equation} \label{eq4}
\underset{k\rightarrow \infty}{\lim}\|\xi_e(k)\|=\underset{k\rightarrow \infty}{\lim}\| \xi_{\text{fb}}(k) - \xi_{r}\|=0
\end{equation}
For the trajectory control problem, the feedback control task is to steer the solution of (\ref{eq3}) to track the time-varying reference such that
\begin{equation} \label{eq5}
\underset{k\rightarrow \infty}{\lim}\|\xi_e(k)\|=\underset{k\rightarrow \infty}{\lim}\| \xi_{\text{fb}}(k) - \xi_{r}(k)\|=0
\end{equation}
%%%%%%%%%%%%%%%%%%%%%%%%%%%%%%%%%%%%%%%%%%%%%%%%%%%%%%%%%%%%%%%%%%%%%%%%%%%%%%%%%%%%%%
\section{NONLINEAR MODEL PREDICTIVE CONTROL DESIGN}
\begin{figure}
\begin{center}
\includegraphics[width=8.4cm]{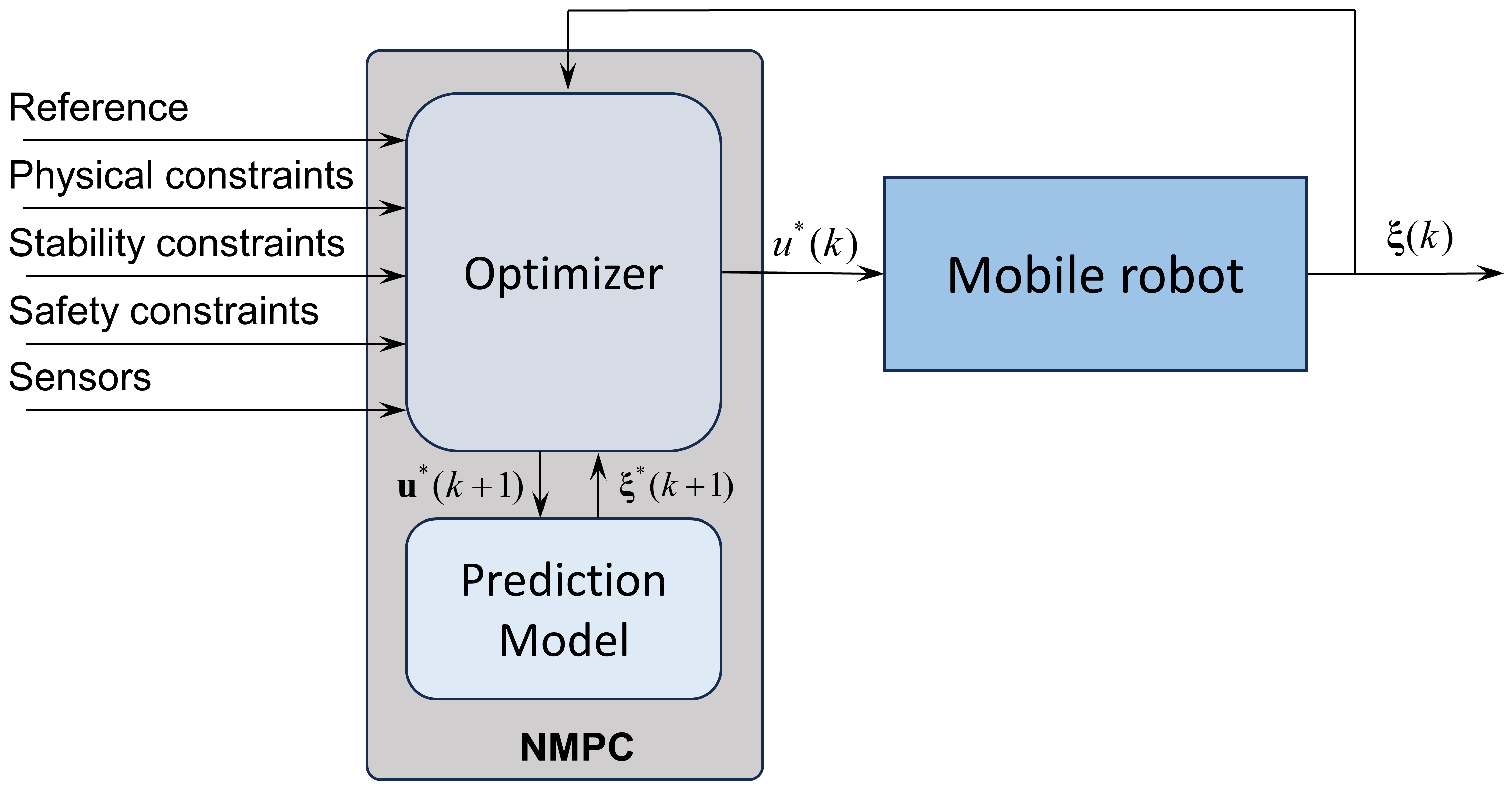}    % The printed column width is 8.4 cm.
\caption{NMPC diagram.} 
\label{Fig2}
\end{center}
\end{figure}

NMPC operates by resolving an open-loop optimisation problem at each measurement interval. In every iteration, the process aims to minimise a predefined cost function. In this section, we design an NMPC framework for the discrete-time system (\ref{eq3}), as depicted in Fig. \ref{Fig2}, integrating CLF and CBF with a relaxed decay rate across the prediction horizon. Some necessary definitions are introduced below.
\\
\subsection{Control Lyapunov Function}
$\textbf{Definition 1.}$ A function $\alpha \,\,:\Re_{\geqslant 0}\rightarrow \Re_{\geqslant 0}$ is of class $\mathcal{K}_{\infty}$-function if it is a $\mathcal{K}$-function and $\alpha(s)\rightarrow +\infty$ as $s \rightarrow +\infty$.
\\
$\textbf{Definition 2.}$ An admissible control invariant set $\mathbf{X}$ for system (\ref{eq3}) is defined as a set where, for all $\xi(t) \in \mathbf{X}$, there exists a state-feedback controller $u(t) \in \mathbf{U}$ such that $\mathbf{F}(\xi(t), u(t))\in \mathbf{X}$. Furthermore, a function $V$ is considered a CLF in $\mathbf{X}$ if, for all $\xi(t) \in \mathbf{X}$, there exist three $\mathcal{K}_{\infty}$-functions $\alpha_1$, $\alpha_2$, and $\alpha_3$ that satisfy $\alpha_1(\|\xi_e(t)\|) \leqslant V(\xi_e(t)) \leqslant \alpha_2(\|\xi_e(t)\|)$ and: 
\begin{equation}\label{eq6}
\varDelta V(\xi_e(t),u(t)) +\alpha_3(\|\xi_e(t)\|) \leqslant 0
\end{equation}
Converting this equation to discrete-time and using $\alpha_1$, $\alpha_2$, $\alpha_3$ as scalars, the CLF constraints are essential for ensuring Lyapunov stability guarantees for any horizon length of NMPC. These CLF constraints establish the existence of a point-wise set of control inputs that achieve stabilisation:
\begin{equation}\label{eq7}
\begin{aligned}
\kappa_{clf} = \{u \in \mathbf{U}:& \Delta V(\xi_e(k), u(k)) \leq -\alpha V(\xi_e(k)),\\
 &0 \leq \alpha \leq 1\}
\end{aligned}
\end{equation}
where $\varDelta V(\xi_e(k),u(k)) = V(\xi_e(k+1))-V(\xi_e(k))$
\\
$\textbf{Remark 1.}$ Refer to \cite{r2.3} for a comprehensive stability analysis of the CLF constraint integrated with NMPC. Furthermore, the research in \cite{r2.4} demonstrates that this method has several desirable properties, such as the absence of a terminal cost and the requirement for fewer tuning parameters. Additionally, \cite{r2.5} successfully applied NMPC with the CLF constraint on a robot platform, resulting in improved computational efficiency.
\subsection{Control Barrier Function}
Inspired by the works done in \cite{r2.6} and \cite{ref26}, this section derives a safety constraint for the optimal control input that guarantees the robot's spatial position, $\xi_{xy}$, always lies within a defined safe set, $\mathscr{Z} \in \Re^{2}$. That is $\mathscr{Z}$ is forward invariant, i.e. if $\xi_{xy}(0) =\xi_{xy0}\in \mathscr{Z}$ then $\xi_{xy}=\xi_{xy}(t) \in \mathscr{Z} ,\forall t$. The set $\mathscr{Z}$ is defined as:
\begin{subequations}\label{eq4}
\begin{align}
&\label{eq18a}\mathscr{Z} =\{\xi_{xy} \in \mathbf{X}\subset \Re^2|B(\xi_{xy}) \geqslant 0 \},\\
&\label{eq18b}\partial \mathscr{Z} =\{\xi_{xy} \in \mathbf{X}\subset \Re^2|B(\xi_{xy}) =0 \},\\
&\label{eq18c}\mathrm{Int}(\mathscr{Z}) \{\xi_{xy} \in \mathbf{X}\subset \Re^2|B(\xi_{xy}) >0 \}.
\end{align}
\end{subequations}
Let $\mathscr{Z}$ denote the superlevel set of a continuously differentiable function $B:\Re^2\rightarrow \Re$. The function $B$ qualifies as a CBF if there exists an $\mathcal{K} _{\infty}$-function $\gamma$ for the control system (\ref{eq3}) to satisfy:
\begin{equation} \label{eq12}
\exists \,\,u(t) \,\,\mathrm{s.t.} \,\,\dot{B}(\xi_{xy}(t),u(t)) \geqslant -\gamma B( \xi_{xy}(t)), \gamma \in \mathcal{K} _{\infty} 
\end{equation}
Extending this inequality constraint to the discrete-time domain and using $\gamma$ as a scalar, we can consider the set consisting of all control values at a point $\xi_{xy}(k) \in \mathbf{X}$ as:
\begin{equation} \label{eq13}
\begin{aligned}
\kappa_{cbf}=\{ u \in \mathbf{U}&: \Delta B(\xi_{xy}(k),u(k)) -\gamma B( \xi_{xy}(k)) \geqslant 0,\\  
&0\leqslant \gamma \leqslant 1\} 
\end{aligned}
\end{equation}
where  $\Delta B(\xi_{xy}(k),u(k))=B(\xi_{xy}(k+1))-B(\xi_{xy}(k))$
$\textbf{Remark 2.}$ By Theorem 2 in \cite{ref26}, if $B$ is a CBF in $\mathbf{X}$ and $\frac{\partial B}{\partial \xi_{xy}}(\xi_{xy}) \ne 0$ for all $\xi_{xy} \in \partial \mathscr{Z}$, then any control signal $u \in \kappa_{cbf}(\xi_{xy}(k))$ for the system (\ref{eq3}) renders the set $\mathscr{Z}$ safe. Additionally, the set $\mathscr{Z}$ is asymptotically stable in $\mathbf{X}$.
\\
The rigid body obstacles are conceptualised as the union of circles with centroids ($x_{ob,i}, y_{ob,i}$) and fixed radius $r_{ob,i}$. Similarly, a circle defines a safe space for the robot with the robot's centre and a radius $r_{rb}$. The CBF is defined for $i$ obstacle as follows:
\begin{equation} \label{eq15}
\begin{aligned}
B_i(\xi_{xy}(k)) =&(x(k) -x_{ob,i}(k)) ^2+( y(k) -y_{ob,i}(k) ) ^2\\&-( r_{rb}+r_{ob,i} ) ^2
\end{aligned}
\end{equation}
In \cite{ref20}, various obstacle avoidance techniques were examined, among which the BT algorithm is also incorporated into NMPC as the safety constraint. This method mandates the robot to navigate tangentially along the obstacle’s perimeter, subsequently reverting to its initial route post-obstacle circumnavigation. The BT algorithm is formulated as follows:
\begin{equation}\label{eq17}
\begin{aligned}
	L_i(\xi_{xy}(k)) =&\sqrt{(x(k)-x_{ob,i}(k) ) ^2+( y(k)-y_{ob,i}(k) ) ^2}\\
	&-( r_{rb}+r_{ob,i} )
\end{aligned}
\end{equation}
\\
Assuming that the robot can detect the position and size of obstacles, represented as $\varphi_{ob} = [x_{ob}, y_{ob}, r_{ob}]^T$, through its sensors at each sampling time, this information is then stored in the parameter vector $p(k)$. This study will demonstrate the improved control performance and safety of CBF utilising the relaxed decay rate technique at a low prediction horizon by comparing Relax-CBF with traditional CBF and BT constraints, as detailed below:
\begin{subequations}\label{eq4}
\begin{align}
&\begin{aligned}
B_i(\xi_{xy}(k+1)&)-(1-\gamma)B_i(\xi_{xy}(k),p(k)) \geqslant 0,\\
&k = 0,...,N-1.
\end{aligned}\\
&L_i(\xi_{xy}(k)) \geqslant 0,\,\, k = 0,...,N.
\end{align}
\end{subequations}
\subsection{NMPC-CLF with Relax-CBF scheme}
The conventional NMPC scheme is formulated by constructing and minimising a cost function and constraints on the prediction horizon grids $N$. The Optimal Control Problem (OCP) is presented in parametric form as follows: 
\begin{subequations}\label{ocp1}
\begin{align}
&\begin{aligned}\mathbf{OCP}\,:\,& J_{N}(\xi (k),u(k),p(k)) = \underset{{\mathbf{S}}}{\min}\underset{m=0}{\overset{N-1}{\sum}}(\varLambda(s_v(m))\\ &+ \varGamma(s_h(m))+\ell( \xi(k+m), u(k+m), p(k))) \end{aligned}\\ 
&\text{Subject to} \nonumber \\ 
&\xi(k) - \xi_{\text{fb}} = 0,\\
&\xi(k+m+1) - \mathbf{F}(\xi(k+m),u(k+m)) = 0,\\
&\xi(k+m) \in \mathbf{X},\,\,u(k+m) \in \mathbf{U},\\
&\begin{aligned}&(1-\alpha)V(\xi(k+m),p(k))+ s_v(m)\\ &-V(\xi(k+m+1),p(k)) \geqslant 0,\,\,\end{aligned}\\
&\begin{aligned}&B_i(\xi_{xy}(k+m+1),p(k))\\ &- s_h(m)(1-\gamma)B_i(\xi_{xy}(k+m),p(k)) \geqslant 0.\end{aligned}
\end{align}
\end{subequations}
where $\mathbf{S}$ contains decision variables as $\mathbf{S}=[\xi(k)^{T},...,\xi(k+N)^{T},u(k)^{T},...,u(k+N-1)^{T}, s_v(k),..., s_v(k+N-1), s_h(k),...,s_h(k+N-1)]$, $p$ is a problem parameter vector that contains reference state vector $\xi_{r}$, feedback state vector $\xi_{\text{fb}}$, and obstacles information, $s_v$ are relaxation terms that avoid conflict between CLF and CBF constraints, and $s_h$ are decay-rate slack variables added to enhance the feasibility of the OCP. Hence, the $\varLambda(s_v)$ and $\varGamma(s_h)$ are included in the cost function to minimise those slack variables. 
\begin{equation}\label{slack}
\begin{aligned}
&\varLambda(s_v(m)) = W_1s_v(m)^2,  \\
&\varGamma(s_h(m)) = W_2(s_h(m)-1)^2.
\end{aligned}
\end{equation}
where $W_1$ and $W_2$ are penalty gains for the slack variables.
\\
The stage cost includes two quadratic terms: one will penalise the tracking error, and another will minimise the control outputs, $\ell:\Re^{3}\times \Re^2\rightarrow \Re_{\geqslant 0}$. The CLF is a quadratic function of the state variable error, $V:\Re^{3}\rightarrow \Re_{\geqslant 0}$.
\begin{equation}\label{eq9}
\begin{aligned}
&\begin{aligned}\ell(\xi(k),u(k),p(k))&=(\xi(k)-\xi_{r}(k))^T W_{\xi} (\xi(k)-\xi_{r}(k))\\ &+(u(k)-u_{r}(k))^TW_{u} (u(k)-u_{r}(k)),\end{aligned}\\
&V(\xi(k),p(k))=(\xi(k)-\xi_{r}(k))^TW_{v}(\xi(k)-\xi_{r}(k)).
\end{aligned}
\end{equation}
where $W_{\xi}$, $W_{u}$, and $W_{v}$ are positive definite symmetric matrices. 
\\
The multiple shooting approach \cite{ref24} is used to transform the OCP into a Nonlinear Program (NLP). The NLP is introduced as below:
\begin{equation} \label{nlp1}
\underset{\mathbf{S}}{\min}\,\,A\left( \mathbf{S} ,p \right) \,\,s.t. \left\{ \begin{aligned}
	&H\left( \mathbf{S} ,p \right) =0,\\
	&G\left( \mathbf{S} ,p \right) \leqslant 0.
\end{aligned} \right. 
\end{equation}
The NLP constraints are divided into two categories: the equality constraint vector, denoted as $H(\mathbf{S}, p)$, ensures that the system dynamics in (\ref{ocp1}b) and (\ref{ocp1}c) are met, and the inequality constraint vector, represented by $G(\mathbf{S}, p)$, imposes restrictions of (\ref{ocp1}d), stability constraints (\ref{ocp1}e) and safety constraints (\ref{ocp1}f) on the decision variables to maintain the system within safe boundaries.
\\
The proposed NMPC strategy is solved using an Interior Point OPTimizer (IPOPT). Based on the current feedback state and environmental data, the IPOPT solver derives an optimal trajectory and control sequence, $\boldsymbol{\xi }_N^*$ and $\boldsymbol{u }_N^*$. The first element of $\boldsymbol{u }_N^*$, labelled $u^*(k)$, is then applied to the system. The residual part of the optimised sequence serves as an initial guess for the decision variable vector in the subsequent iteration. Thereafter, the solver is re-engaged to deduce new control input values.
\\
$\textbf{Remark 3.}$ As analysed in \cite{r2.8}, the proposed control schemes with relaxed CLF and CBF constraints consistently maintain optimisation feasibility. Furthermore, with the implementation of the decay-rate relaxation technique for CBF, as introduced by \cite{r2.9}, the safety performance of agents is enhanced by reducing $\gamma$ while not harming the feasibility of the OCP.
%%%%%%%%%%%%%%%%%%%%%%%%%%%%%%%%%%%%%%%%%%%%%%%%%%%%%%%%%%%%%%%%%%%
\section{RESULTS AND DISCUSSION}
This section provides numerical and experimental validation of the proposed scheme on a mobile robot that operated in a dynamic environment containing random static and moving obstacles.
\subsection{Point Stabilization with Obstacle Avoidance}
Consider the problem of regulating the system (\ref{eq3}) to a target state while ensuring safety within the context of set invariance. The robot was controlled from an initial pose $\xi_0=\left[0,0,0 \right]^T$ (m,m,rad) to reach a target pose $\xi_{r}=[10, 0, 0]^T$ (m,m,rad), encountering a static obstacle at locations (0.1m, 4.7m) with radius 0.6 m. The robot's radius was 0.25 m. The sampling time was $\tau = 0.1$ s, and the prediction horizon was $N = 5$. The control input constraints were set as $[-1\,\, -\pi/4]^T\leqslant [v\,\,\, w]^T \leqslant [1\,\, \pi/4]^T$. The control parameters were as follows: $W_{\xi} = \text{diag}(10, 15, 0.7)$, $W_u = \text{diag}(1, 0.1)$, $W_{v} = \text{diag}(1, 1, 1)$, $W_1 = 70$, $W_2 = 10$, and $\alpha = 0.3$.
\begin{figure}[h!]
\centering
\includegraphics[width=7cm]{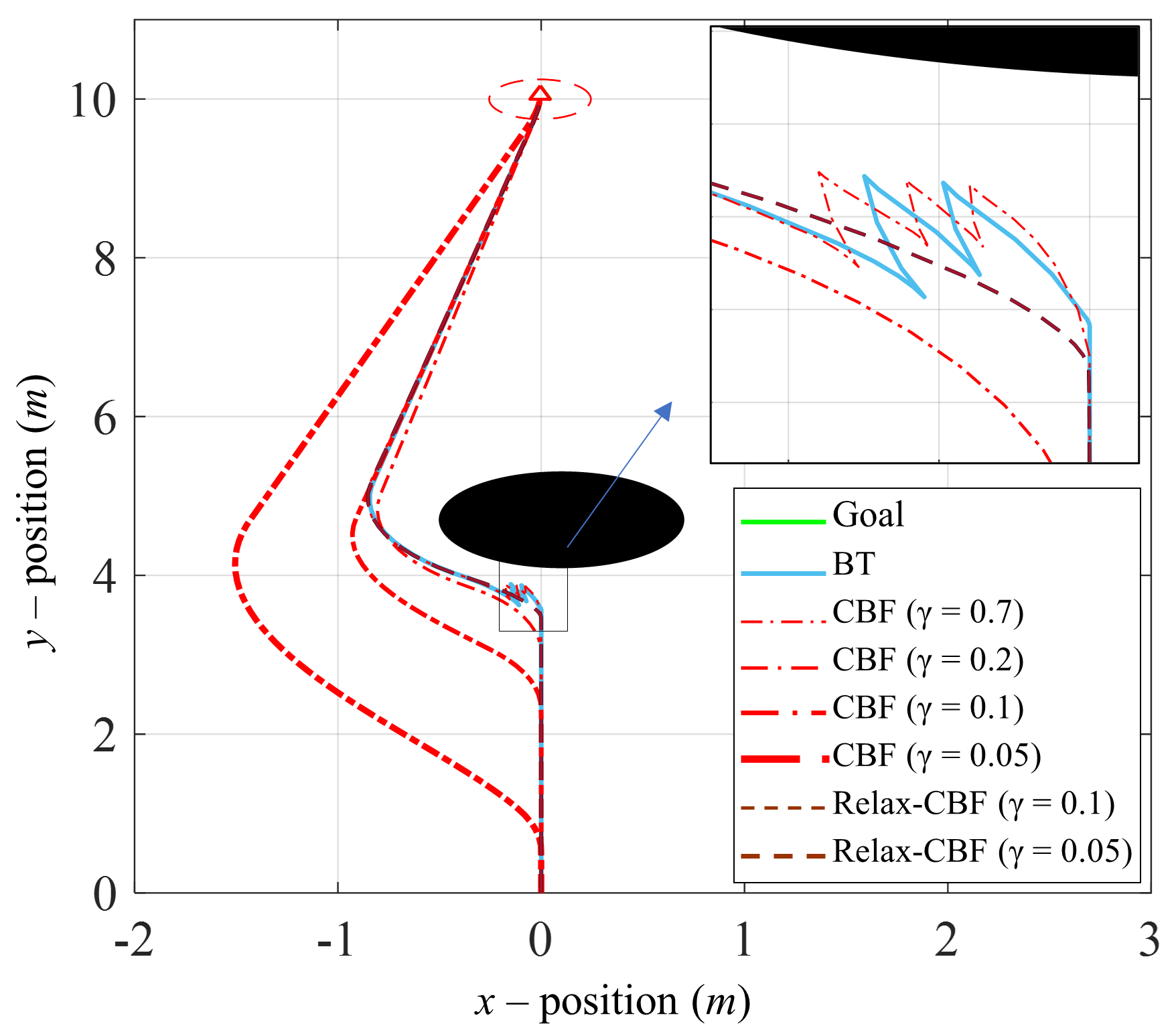}
\caption{Point stabilization with obstacles using $N = 5$ and different $\gamma$ values.}
\label{ps_m}
\end{figure}
\begin{figure}[h!]
\centering
\includegraphics[width=7cm]{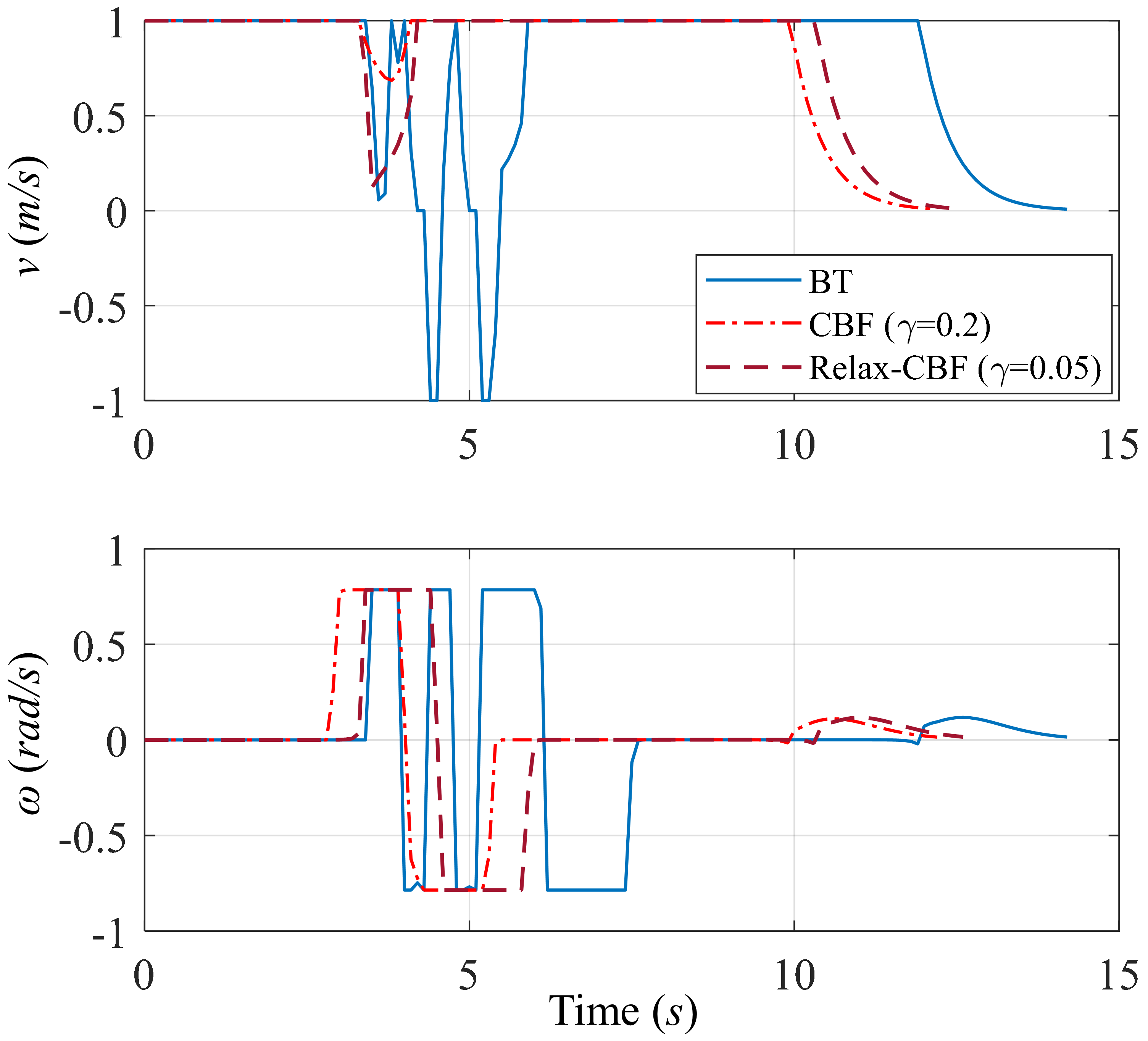}
\caption{Crossvalidated control signal for point stabilisation with obstacles.}
\label{ps_ct_m}
\end{figure}
\\
Point stabilisation results were depicted in Fig. \ref{ps_m}. We illustrated the robot's trajectory using various colours to represent different choices of $\gamma$, and a black circle represented an obstacle. With a prolonged prediction horizon ($N \geqslant 12$), both the CBF and BT methods had successfully navigated the robot to the target point without compromising safety. Notably, CBF had surpassed BT in scenarios with shorter prediction horizons because CBF restricted the robot's movements, enabling early action at a substantial distance from obstacles. In contrast, the BT algorithm proved effective only at close range, often necessitating frequent stops and direction reversals to prevent collisions. This behaviour was evidenced by the negative values in the linear velocity of the BT controller, as shown in Fig. \ref{ps_ct_m}. Therefore, the advantage of integrating CBF with NMPC lies in ensuring a smooth trajectory with a shorter horizon, thereby reducing the computational burden. However, setting $\gamma$ too low for traditional CBF significantly reduced the reachable set of the controller, leading to a trade-off between feasibility and safety. As demonstrated with $\gamma = 0.05$, Relax-CBF showcased its superiority by achieving optimal control performance without affecting feasibility.
\begin{figure}[h!]
\centering
\includegraphics[width=8.5cm]{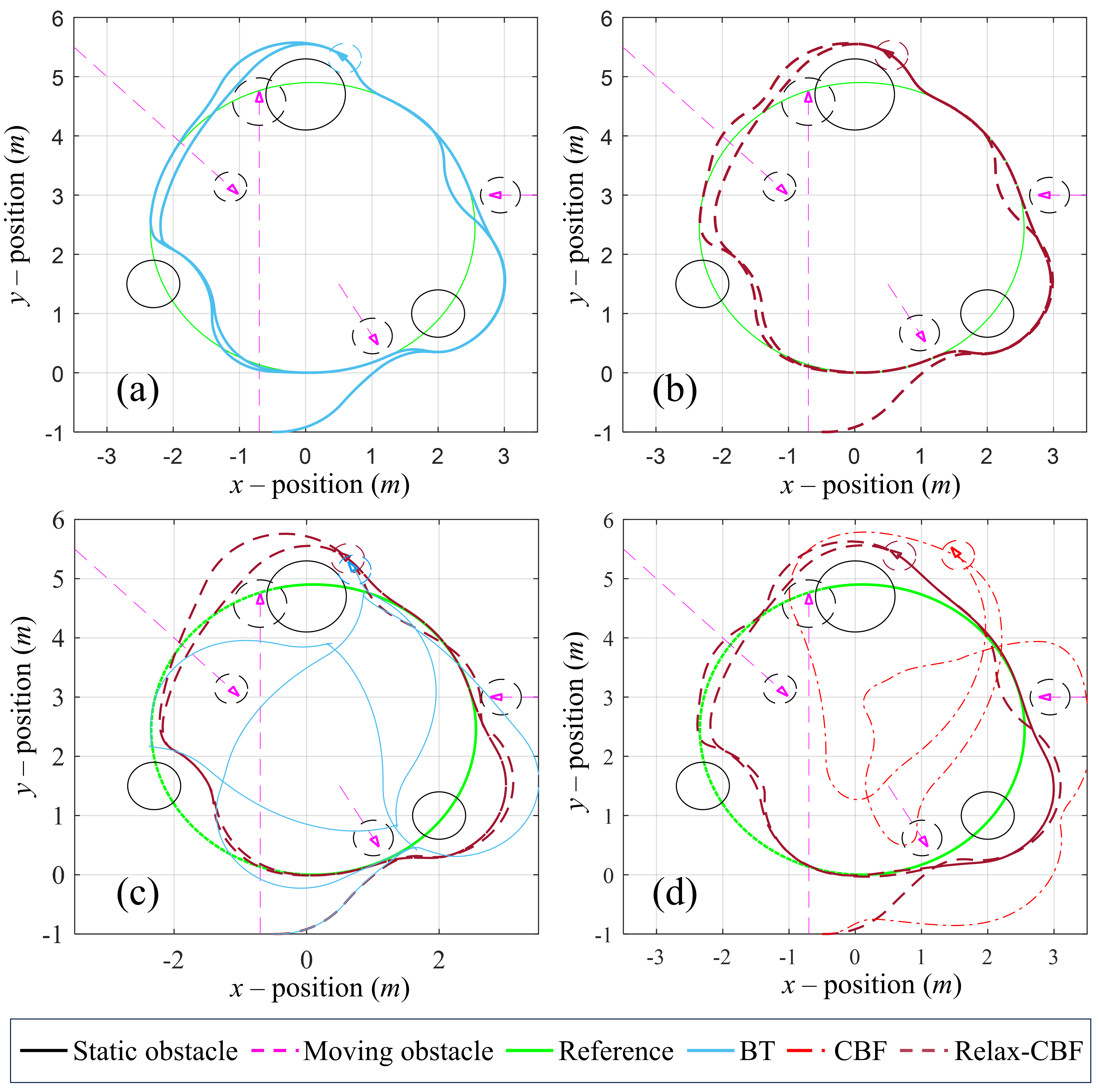}
\caption{Crossvalidated control performance for trajectory tracking in a dynamic environment: (a) BT ($N=12$); (b) Relax-CBF ($N=12,\gamma=0.2$); (c) BT ($N=5$) and Relax-CBF ($N=5,\gamma=0.12$); (d) CBF ($N=5,\gamma=0.03$) and Relax-CBF ($N=5,\gamma=0.03$).}
\label{ct_m}
\end{figure}
\begin{figure}[h!]
\centering
\includegraphics[width=7cm]{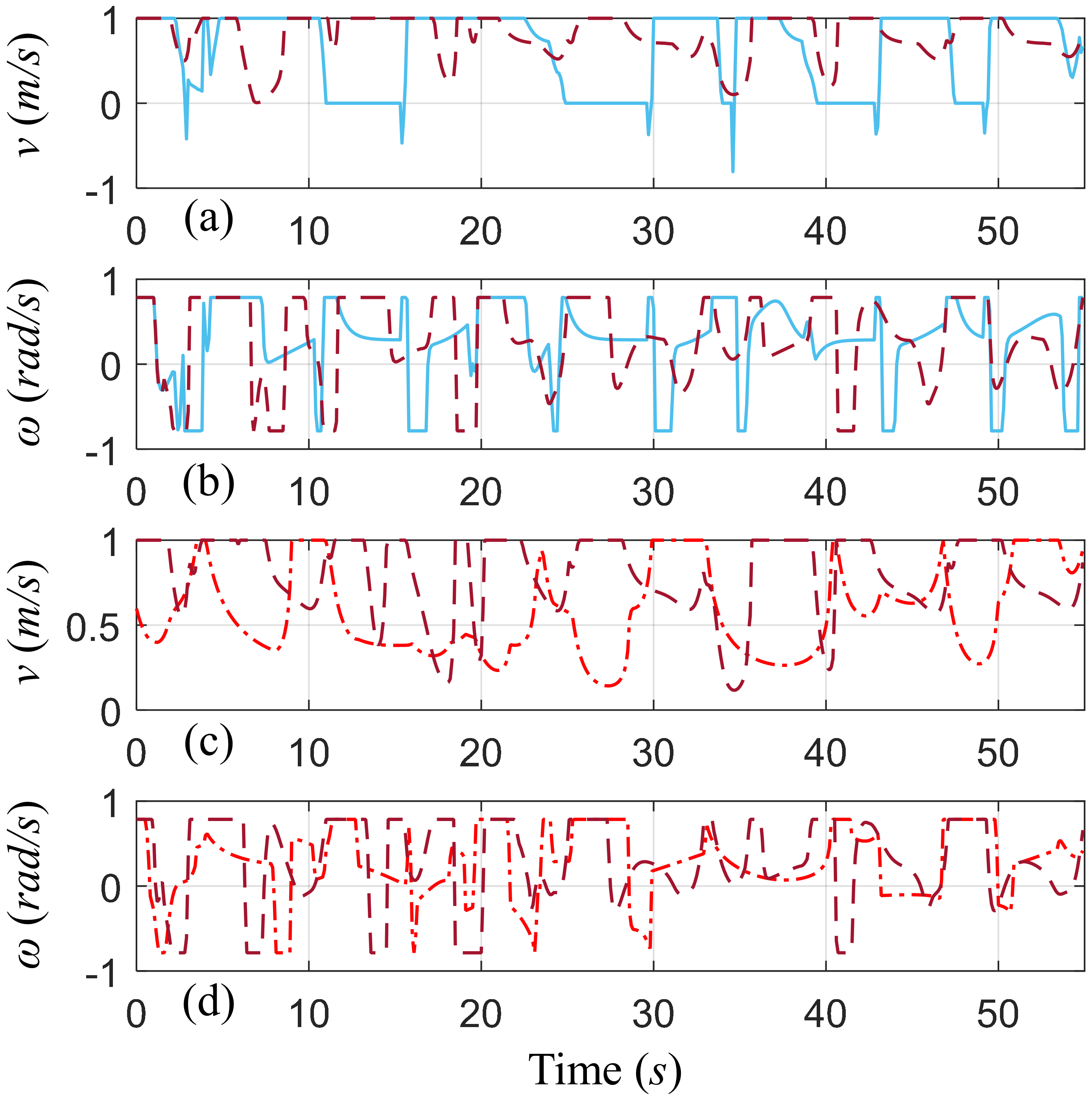}
\caption{Trajectory tracking control signal. (a), (b) for simulations in Fig. 4 (c) and (c), (d) for simulations in Fig. 4 (d).}
\label{ct_ct_m}
\end{figure}
\subsection{Trajectory tracking in dynamic environment}
In the second simulation, trajectory tracking tasks were conducted, with a reference trajectory following a circular path defined as $[x_r(t)=0.7\cos(0.03t), \,\, y_r(t) = 0.7\sin(0.03t)]^T$. 
The test included three static obstacles at (2m, 1m), (0m, 4.7m), and (-2.3m, 1.5m) with r = 0.4 m, r = 0.6 m, and r = 0.4 m, and four moving obstacles with initial pose, linear velocity, and radius $(-8, 10, -0.78, 0.63, 0.25)$, $(0.5, 1.5, -1, 0.4, 0.3)$, $(-0.7, -5, 1.5, 0.31, 0.4)$, and $(5.5, 3, -3.14, 0.34, 0.3)$. Control parameters and the update rate were the same as in the previous simulation. Fig. \ref{ct_m} (a) and (b) depict the safe trajectories in the $xy$-plane regulated by BT and Relax-CBF at $N=12$, with moving obstacles marked by pink dashed lines. Both controllers exhibited efficient tracking performance and obstacle avoidance. However, for the short prediction horizon of $N = 5$, Fig. \ref{ct_m} (c) shows that Relax-CBF initiated earlier deviations around obstacles, leading to smoother navigation. In contrast, the BT algorithm often approached the boundaries of obstacles closely, leaving the robot stuck there for periods of time. Regarding safety, due to the use of low $\gamma$, the traditional CBF was unable to track the reference as the feasible state region was dramatically reduced. Meanwhile, the feasibility of Relax-CBF was independent of $\gamma$. As a result, Relax-CBF still guaranteed efficient tracking performance in Fig. \ref{ct_m} (d). Control signals remained within their limitations in Fig. \ref{ct_ct_m}.
\begin{figure}[h!]
\centering
\includegraphics[width=6.5cm]{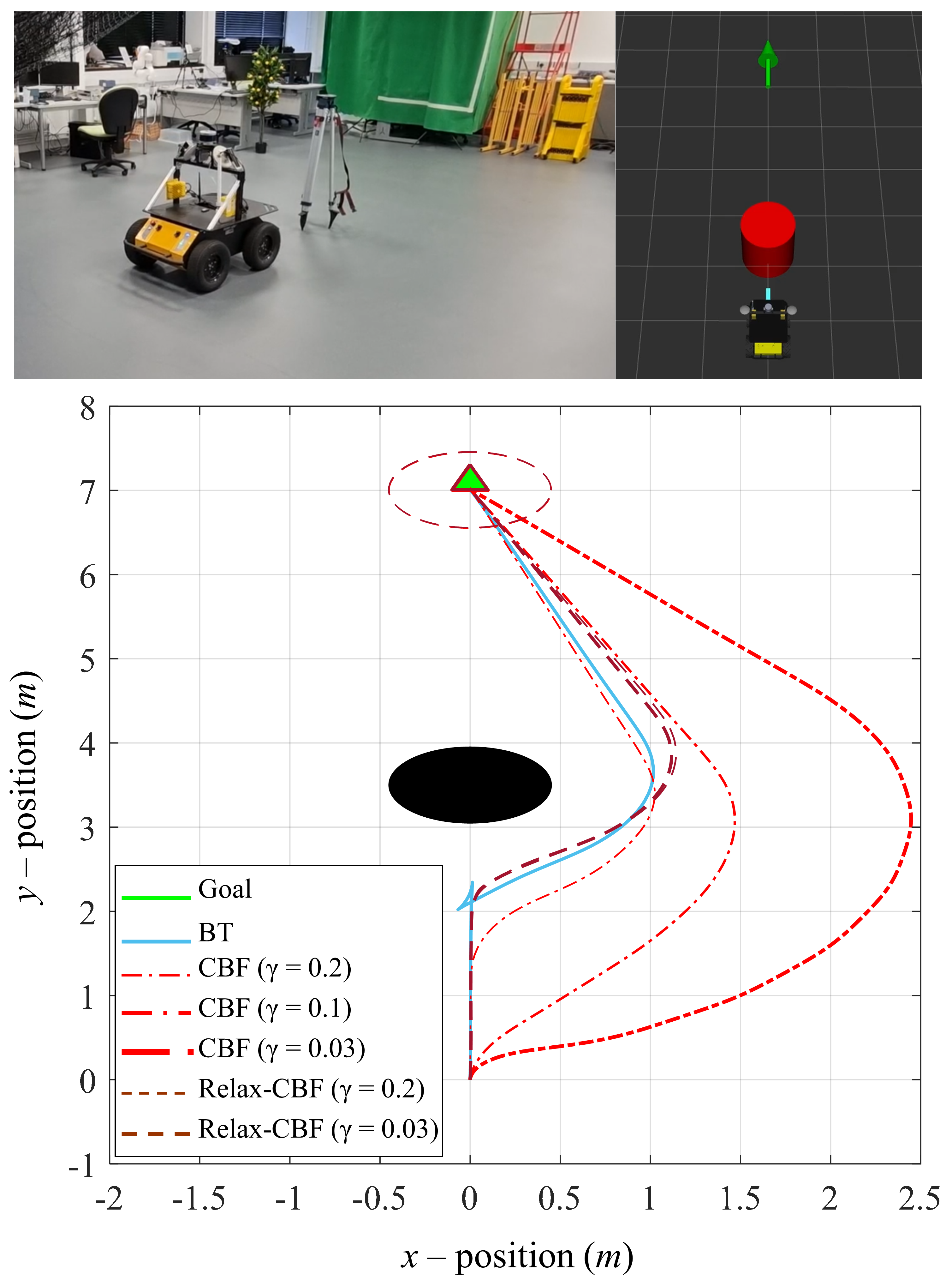}
\caption{Crossvalidated point stabilisation on Husky A200}
\label{ex_ps}
\end{figure}
\begin{figure}[h!]
\centering
\includegraphics[width=6.5cm]{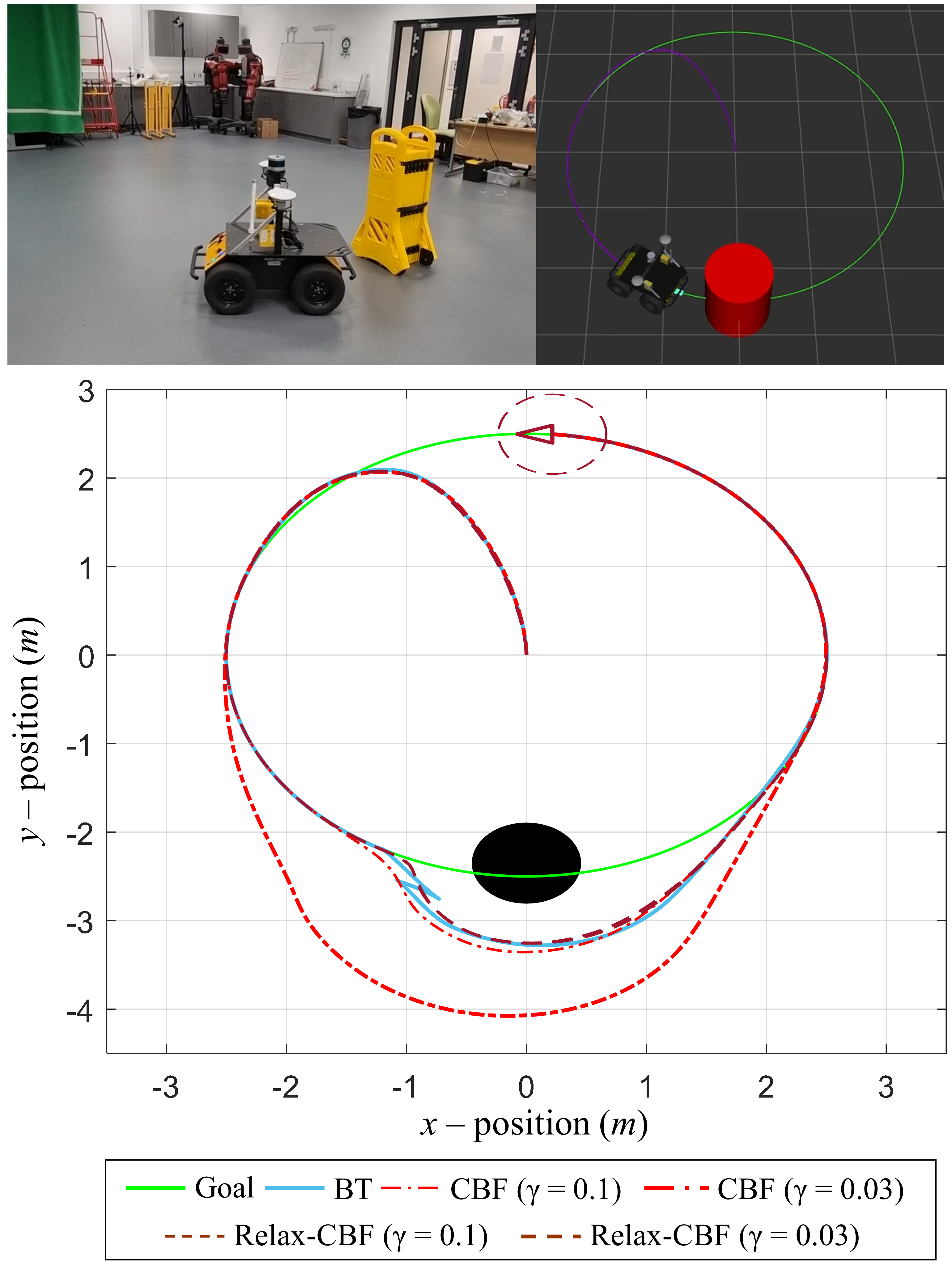}
\caption{Crossvalidated trajectory tracking on Husky A200 
			\\(See Video 1 at \href{https://youtu.be/fUrT08NPNrA}{https://youtu.be/fUrT08NPNrA}).}
\label{ex_ct}
\end{figure}
\subsection{Experimental Validation of Controller Performance}
In this section, we conducted experimental validations of the proposed NMPC-CLF-Relax-CBF and other safety-critical controllers using the Husky A200 robot, as depicted in Fig. \ref{ex_ps} and Fig. \ref{ex_ct}. A detailed video of the experiment is available online (Video 1). The Husky was programmed to move from $\xi_0=[0,0,0]^T$ (m,m,rad) to $\xi_r=[7,0,0]^T$ (m,m,rad), navigating around an obstacle located at (3.5m, 0m) with a radius of 0.45 m. Additionally, a circular path was defined as $[x_r(t)=2.5 \cos(0.16t), , y_r(t) = 2.5 \sin(0.16t)]^T$ with an obstacle $r_{ob}=0.45$ m positioned at (0m, -2.35m). The control parameters were set as follows: $W_{\xi} = \text{diag}(15, 15, 0.1)$, $W_u = \text{diag}(0.01, 0.01)$, $W_{v} = \text{diag}(1, 1, 1)$, $W_1 = 100$, $W_2 = 10$, and $\alpha = 0.2$. The chosen sampling time was $\tau = 0.2$ s, with $N = 5$ for the point stabilisation problem and $N = 7$ for circle tracking. The experimental results were consistent with the outcomes obtained from the MATLAB simulations, demonstrating that all controllers achieved effective target positioning and reference tracking. The BT controller allowed the robot to approach the obstacle more closely, necessitating stops and direction reversals to navigate around it. The CBF controller generated a better strategy for obstacle avoidance; however, it could only achieve smooth and optimal trajectories with appropriate values of $\gamma$. The Relax-CBF demonstrated that the proposed NMPC-CLF-Relax-CBF effectively ensures the robot's safe and accurate navigation with short prediction times and a significantly wide range of $\gamma$ values, simultaneously enhancing feasibility, safety, and computational efficiency. By employing Relax-CBF, the NMPC was able to reduce the prediction time ($N\tau$) from 1.2 seconds to 0.5 seconds, achieving a 58.33\% reduction in simulations, and from 3 seconds to 1.4 seconds, achieving a 53.33\% reduction in experiments.
\section{CONCLUSIONS}
This paper presented a formulation for a safety-critical nonlinear model predictive control scheme that incorporates a CLF for system stability and Relax-CBF to impose safety constraints. We simulated three controller schemes to achieve two common control objectives for mobile robots: point stabilisation and trajectory tracking with static and dynamic obstacles. Various challenging scenarios were designed to highlight the effectiveness of each control scheme. The results demonstrated that all schemes can achieve reliable and accurate performance. However, the CBF schemes successfully operated with a smaller prediction horizon, making them more computationally efficient for practical implementation. In terms of safety, the CBF with a relaxed decay rate demonstrated a promising solution to resolve the trade-off between feasibility and safety in optimal control problems. Experimental validations have confirmed the effectiveness of the proposed controllers in real-world settings. Future work will adapt this controller to experimental scenarios with moving obstacles and multi-agent systems.
%\section*{APPENDIX}
%\section*{ACKNOWLEDGMENT}

\end{document}